\documentclass[pra,preprint,showpacs,floatfix]{revtex4}
\usepackage{graphicx}
\usepackage{psfrag}
\usepackage{amssymb,amsmath}
\begin{document}

\title{Radius of a Photon Beam with Orbital Angular Momentum}

\author{B.~S.~Davis}

\email{bdavis2@tulane.edu}

\affiliation{Department of Physics, Tulane University, New Orleans, Louisiana 70118, USA}

\author{L.~Kaplan}

\email{lkaplan@tulane.edu}

\affiliation{Department of Physics, Tulane University, New Orleans, Louisiana 70118, USA}

\begin{abstract}
We analyze the transverse structure of the Gouy phase shift in light beams carrying orbital angular momentum and show that the Gouy radius $r_G$ characterizing the
transverse structure grows as $\sqrt{2p+|\ell|+1}$ with the nodal number $p$ and photon angular momentum number $\ell$. The Gouy radius is shown to be closely related to the root-mean-square radius of the beam, and the divergence of the radius away from the focal plane is determined. Finally, we analyze the rotation of the Poynting vector in the context of the Gouy radius.
\end{abstract}

\date{\today}

\maketitle 

\section{Introduction}

Light beams carrying orbital angular momentum (OAM)~\cite{PoyntingOAM,Beth,Barnett}, which is distinct from spin (or polarization) angular momentum, have become important subjects for theoretical and experimental investigation over the past two decades,
and have been studied in the context of applications as diverse as angular momentum transfer to trapped particles~\cite{he,SimpsonDAP}, quantum communication~\cite{gibson,barreiro,WildeUskov1}, and sensing~\cite{Foo}. 
The OAM is directly related to the azimuthal phase of the mode, independent of the axis about which the phase angle is defined~\cite{Berry}. Photons exhibiting orbital angular momentum are also referred to as twisted photons \cite{Molina-TerrizaTorresTorner,BabikerBADR}. A collimated coherent beam of such photons is called an optical vortex~\cite{MolinaTerrizaRTTW}, a twisted beam~\cite{BerryMcDonald}, or a helical laser beam~\cite{GiovanniniNMS}.
The interaction of beams carrying OAM with matter has become a subject of considerable interest~\cite{BabikerBADR,PadgettAllen,AlexandrescuCDiF,SimulaNHCSM,SimpsonDAP,BabikerPA}.

In the paraxial approximation, cylindrically symmetric light beams carrying OAM may be described by Laguerre-Gaussian (LG) modes, which are a natural generalization of the pure Gaussian beam without OAM.
Laser beams with a pure Gaussian transverse profile are characterized by a Gouy phase of the form $$\exp[-i\theta_{0,0}]=\exp [-i\tan^{-1} (z/z_R)]\,,$$ where $z$ is along the direction of propagation, $z_R$ is the Rayleigh range (to be defined below), and the focal point or beam waist is taken to be at $z=0.$ It is evident that as the beam propagates from $-\infty$ to $+\infty$ it undergoes a phase shift of $\pi$ relative to a plane wave~\cite{Gouy}. A modified Gouy phase appears for an LG beam with winding number $\ell$ and radial node number $p \ge 0$,
$$\exp[-i\theta_{p,\ell}]= \exp\left [ -i(2p +|\ell| +1)\tan^{-1} \left ( \frac{z}{z_R} \right )\right ]\,.$$
Thus the corresponding cumulative Gouy phase change of an LG beam is $\pi(2p+|\ell|+1).$

Historically the Gouy effect has been viewed primarily -- if not exclusively -- as a longitudinal phenomenon, with the beam accumulating a phase change of $\pi$, or more generally $\pi(2p + |\ell| +1)$, as it passes along the symmetry axis through the focal plane \cite{HariharanRobinson,PadgettAllen,McGowanCG,BerryMcDonald}. However, there is also a transverse Gouy effect that has received scant attention.

We will show that the Gouy phase shift in an LG beam has a transverse radial dependence. The Gouy effect is maximal along the axis of propagation, conventionally taken to be the $z$ axis, and decreases radially outwards. At a specific value of the radial coordinate $r$, the Gouy phase change vanishes, and it becomes negative for higher values of $r$. We shall call this critical value of $r$ the \textit{Gouy radius} and label it as $r_G$. 

In what follows we shall establish that:
(1) In the focal plane of a pure Gaussian beam, the Gouy radius $r_G$ equals the beam waist $w(0)$. 
(2) In the focal plane of an LG beam, the Gouy radius $r_G$ becomes $w(0)\sqrt{2p+|\ell|+1}$, and is proportional to the rms radius of the beam, which also
grows as $\sqrt{2p+|\ell|+1}$.
(3) The Gouy radius $r_G$ increases with $z$, becoming divergent at the Rayleigh range $z_R$, even while the width $w(z)$ and the rms radius $r_{\rm rms}(z)$ remain finite.
(4) The Gouy radius serves as a natural physical boundary for the photon beam, delimiting the ``allowed'' zone from a ``forbidden'' zone.
(5) The Gouy radius is helpful in understanding the rotation of the Poynting vector in a twisted photon beam.

\section{Laguerre Gaussian Beams}
\label{lgbeams}

A beam of photons propagating in the $z$ direction prepared with linear momentum $k_0$, orbital angular momentum $\ell$, and radial node number $p$ may be represented by the vector potential~\cite{AllenBSW}
\begin{equation}
\mathbf{A} = \hat{\mathbf{x}} \,u(r, z,\phi)\,e^{ik_0 z}\label{A}\,,\end{equation} where $\hat{\mathbf{x}}$ is a unit vector in the $x$ direction and \cite{AllenPadgettBabiker,YaoPadgett}:
\begin{eqnarray}
u (r,z,\phi) &=&LG_{p,\ell} (r,z,\phi)= \sqrt{\frac{2p!}{\pi (p+|\ell|)!}} \, \frac{1}{w(z)} \left [ \frac{\sqrt{2}r}{w(z)}\right ]^{|\ell|}\exp\left [ -\frac{r^2}{w^2(z)} \right ]L_{p}^{|\ell|}\left ( \frac{2r^2}{w^2(z)} \right )\nonumber \\
&\times & \exp[i\ell\phi]\,\exp\left [ \frac{ik_0r^2z}{2(z^2+z_R^2)} \right ] \exp\left [ -i(2p +|\ell| +1)\tan^{-1} \left ( \frac{z}{z_R} \right )\right ]\label{LG}\,,\end{eqnarray}  
where $L_{p}^{|\ell|}$ are the generalized Laguerre polynomials, and the Rayleigh range $z_R$ will be defined below.
For a pure Gaussian beam ($p=\ell=0$), the radial dependence of the amplitude is given simply by
\begin{equation} u (r,z,\phi) \sim \exp \left[-\frac{r^2}{w^2(z)}\right]\,,\label{gbeam}\end{equation}
and the beam width or radius of such a beam is conventionally defined as the radial distance at which the vector potential falls to $1/e$ of its value on the central axis $r=0$. Thus by definition the width or radius of a Gaussian beam is $w(z)$.

A general LG beam, however, has an additional radial structure, including $p$ radial nodes.  Yet the parameter $w(z)$ that appears in Eq. (\ref{LG}) is still conventionally referred to as the width or radius of the LG beam~\cite{AllenBSW,CourtialPadgett,PadgettAllen,LuoRSW,BerryMcDonald,TabosaPetrov}. But this is misleading for nonzero $\ell$ or $p$, as we shall demonstrate below. 

At the beam waist of a Gaussian beam, where the radius is $w(0)$, the cross sectional area is $A = \pi w^2(0)$.
$A$ is also the area of the aperture from which the Gaussian beam emerges.
A quantity of importance for laser beams is the Rayleigh range $z_R \equiv A/\lambda$,
where $\lambda$ is the wavelength of the beam (Ref.~\cite{Siegman}, p. 714).
So an important relationship exists between the beam waist radius $w(0)$, the Rayleigh range $z_R$ and the wavelength $\lambda$ of a Gaussian beam \cite{Siegman}:
\begin{equation}  w^2(0) = \frac{\lambda z_R}{\pi}\,.\label{Gauss}\end{equation}
Equation~\ref{Gauss} may be used as an alternative definition of the Rayleigh range for a Gaussian beam. The Rayleigh range sets the scale on which the Gaussian beam spreads~\cite{Siegman},
\begin{equation} w(z) = w(0)\left[\frac{z^2 +z^2_R}{z^2_R}\right]^{1/2}\,. \label{waist1}\end{equation}

\begin{figure}
	\centering
		\includegraphics[width=4in]{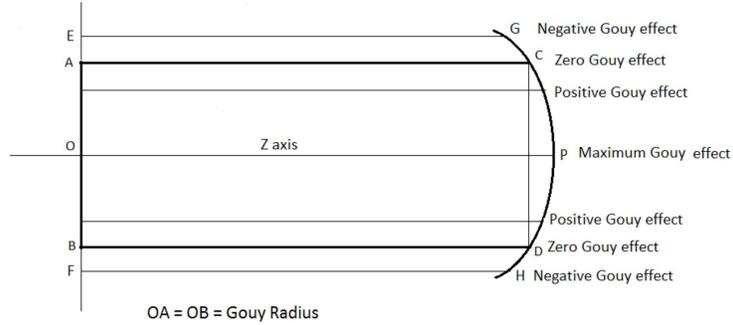}
		\caption {Section of a Gaussian Beam.}
\label{figbeam} 
\end{figure}
\section{Transverse Gouy Effect for Gaussian and LG Beams} 

Because of the divergence of the Gaussian beam as it propagates away from the beam waist at $z=0$, the wavefront is convex rather than planar, as illustrated in Fig.~\ref{figbeam}. So the longitudinal wave vector component $k_z$ is smallest along the $z$ axis (where the Gouy effect is strongest and the phase advances most slowly) and increases with the distance from this axis. Of course, for a perfectly collimated beam with $z_R = \infty $, the wave front would be planar, and $k_z$ would be everywhere a constant equal to $k_0$. But for a Gaussian beam $k_z$ is not constant across the transverse profile. One might expect that $k_0$ would equal the value of $k_z$ at the center of the beam, on the $z$ axis. But this is not the case. Because of the Gouy effect, the value of $k_z$ on the $z$ axis is necessarily smaller than the wave number $k_0$ which would correspond to a plane wave front.

In what follows we shall examine the Gaussian beam at its waist, at $z=0$, and show that it is precisely on the circumference of the beam waist, at $r=w(0)$, that $k_z = k_0$, and that $k_z$ increases monotonically with $r$.

The normalized field pattern for a Gaussian beam is given by setting $p=\ell=0$ in the general expression of Eq.~(\ref{LG}) for LG beams:
\begin{equation}
u(r,z,\phi) = \sqrt{\frac{2}{\pi}} \,\frac{\exp[ik_0z -i\psi(z)]}{w(z)} \exp \left[-\frac{r^2 }{w^2(z)} +\frac{ik_0r^2z}{2(z^2+z_R^2)}\right]\,, \label{Gaussianwaist}\end{equation}
where $ \psi (z) = \tan^{-1}(z/z_R) $. Now
$k_z$ can be evaluated as a function of the radial distance $r$ at the beam waist $z=0$, by obtaining the derivative $-i\partial u/\partial z$.
To first order in $z$, $\psi(z)\rightarrow z/z_R$,  and from Eq.~(\ref{waist1}) $w(z)\rightarrow w(0)$,
so
\begin{equation}
k_z(r)_{z=0} =k_0 - \frac{1}{z_R} + \frac{k_0 r^2}{2z^2_R} \,. \label{Gaussmomentum} \end{equation}
It is evident that $k_z = k_0$ when
$$ r^2 = \frac{2z_R}{k_0}= \frac{z_R \lambda}{\pi}\,,$$
where $\lambda= 2\pi/k_0$ is the photon wavelength.
Clearly, the Gouy radius at $z=0$ is $r_{G}(0) = w(0).$ So we have established the important result that for a Gaussian wave at its waist or focus, the Gouy radius is identical with the beam waist radius $w(0)$.

Similarly for the general LG beam (Eq.~(\ref{LG})), we have
\begin{equation} k_z(r)_{z=0} = k_0 +\frac{k_0 r^2 -2(2p +|\ell| +1)z_R}{2z^2_R}  \,. \label{zmomentum}\end{equation}
This yields the Gouy radius
\begin{equation} r^2_{G} (0) = (2p + |\ell| +1)\frac{\lambda z_R}{\pi}\label{waist2}=(2p+|\ell|+1) w^2(0)\,.\end{equation}

\section{Radial Intensity Profile of Twisted Beams}

The amplitude of the LG mode is given by 
$$|u(r,z,\phi)|=\sqrt{\frac{2p!}{\pi (p+|\ell|)!w^2(z)}}x^{|\ell|/2} e^{-x/2} L^{|\ell|}_p(x)\,, $$
where $x = \frac{2r^2}{w^{2}(z)}$.
This function oscillates in a range $x_{-} < x < x_{+}$, where
\begin{equation}
x_{\pm} = 2p +|\ell| + 1 \pm \sqrt{(2p+|\ell| + 1)^2 - (|\ell| -1)^2} \end{equation}
and decays exponentially outside this range~\cite{BerryMcDonald}. We then have
\begin{equation} \langle r^2 \rangle =
\int_{0}^{\infty } 2\pi r^3 dr |u (r,z,\phi)|^2 = \frac{w^2(z)}{2}(2p + |\ell| + 1) \,,\label{rsquared}\end{equation}
and
\begin{equation} r_{\rm rms}(0) = w(z) \sqrt{\frac{2p+|\ell|+1}{2}}\,. \end{equation}
Thus for a pure Gaussian wave, with $p=\ell=0$, $r_{\rm rms} = w(z)/\sqrt{2}$. Notably, we find that at the beam waist $z=0$, the rms beam radius is $1/\sqrt{2}$ times the Gouy radius computed in Eq.~(\ref{waist2}), for all 
values of $p$ and $|\ell|$.

Clearly, $r_{\rm rms}$ is not the same as the radial distance of maximum intensity $r_{\rm max}$ which has the value (for $p=0$) \cite{PadgAll}
\begin{equation} r_{\rm max} (z) = w(z)\sqrt{\frac{|\ell|}{2}}\,.\label{rmax} \end{equation}
The radius of maximum intensity becomes zero for a pure Gaussian wave ($p =\ell = 0$), but the expected or rms value of $r$ does not vanish for any realistic distribution. For large values of $\ell$ at fixed $p$, $r_{\rm max}$ approaches $r_{\rm rms}$.

\section{Beam width away from the waist}
\label{away}
We shall now tackle the more general problem of finding the beam width of a twisted photon beam for arbitrary values of $z$, not just at the beam waist or focus. A note of caution is in order, because Eq.~(\ref{LG}) is a solution to the Maxwell equations only in the paraxial approximation~\cite{AllenBSW}. So for a divergent beam the equation is physically meaningful only for values of $z$ sufficiently small in relation to the Rayleigh range. 
We will write Eq.~(\ref{LG}) in compact form as
\begin{equation}
u = F(r,z) \exp iG(r,z,\phi)\,,\label{rweqn1}\end{equation}
where $F$ and $G$ are real. So using Eq. (\ref{A}) we may write
\begin{equation} k_z = k_0 - \frac{i}{u}\frac{\partial u}{\partial z} \end{equation}
 As before, we will obtain the wave vector or momentum component $k_z$, this time as a function of both $r$ and $z$:
\begin{equation}  -i\frac{\partial u}{\partial z} = -i\frac{1}{F}\frac{\partial F}{\partial z}u +\frac{\partial G}{\partial z} u \,. \label{rweqn2}\end{equation}
The photon momentum is the sum of an imaginary and a real term. The former corresponds to an amplitude change, and the latter to a phase change~\cite{HariharanRobinson}.  We are interested in the phase change, so
\begin{equation} 
k_z = k_0 +\frac{\partial G}{\partial z} = k_0+\frac{k_0r^2(z^2_R -z^2) - 2z_R(z^2_R +z^2)(2p+|\ell|+1)}{2(z^2_R +z^2)^2} \,. \label{kzaway} \end{equation}
Using the definition of the Gouy radius $r_{G} (z)$ as the value of $r$ for which this change in phase becomes zero, we have
\begin{equation}
r^2_G(z) = \frac{\lambda z_R}{\pi} (2p + |\ell| +1)\frac{(z^2_R + z^2)}{(z^2_R -z^2)}
=w^2(0) (2p + |\ell| +1)\frac{(z^2_R + z^2)}{(z^2_R -z^2)}\,.
\label{GRz1} \end{equation}
This expression reduces to Eq.~(\ref{waist2}) as $z\rightarrow 0$. The Gouy radius also diverges as $z\rightarrow z_R$. But as stated earlier, this latter case is not physical, as the paraxial approximation of Eq.~(\ref{LG}) breaks down for large values of $z$. For $|z|\ll z_R$ we may simplify the result as
\begin{equation}r^2_{G} (z) =w^2(0)(2p + |\ell| +1)\left(1+\frac{2z^2}{z^2_R}\right)\,. \label{GRz2} \end{equation}

\section{Total Linear Momentum of a Laguerre Gaussian Beam}
\label{linmomentum}
The linear momentum of a Gaussian wave photon differs from the corresponding momentum of a plane wave photon because of the Gouy phase shift ~\cite{HariharanRobinson}, as is evident from Eq.~(\ref{zmomentum}). We will now calculate the linear momentum of a Laguerre Gaussian photon.

We first find the average momentum density integrated over the beam waist, i.e., over the $z=0$ focal plane.
From Eq.~(\ref{zmomentum}) we obtain
\begin {equation} 
\langle k_z \rangle_{z=0} = k_0 +\frac{k_0}{2z^2_R} \langle r^2 \rangle - \frac{2p+|\ell| +1}{z_R}\,,\end{equation}
and using Eq.~(\ref{rsquared}), we have
\begin{equation} \langle k_z \rangle_{z=0} = k_0 -\frac{2p+|\ell| +1}{2z_R}\,.\label{kzLG}\end{equation}
For a pure Gaussian beam, this reduces to 
$\langle k_z \rangle_{z=0} = k_0 -1/2z_R$.
For a plane wave, for which $z_R=\infty$, the photon momentum reduces to $k_0$.

To obtain the average momentum density at arbitrary $z$, we use Eq.~(\ref{kzaway}) and average over a plane normal to the $z$ axis (the propagation direction):
\begin{equation}
\langle k_z \rangle = k_0 + \frac{k_0(z^2_R -z^2)}{2(z^2_R + z^2)^2} \langle r^2 \rangle - \frac{(2p+|\ell| +1)z_R}{z^2 +z^2_R} =  k_0 -\frac{(2p+|\ell|+1)}{2z_R} \,. \label{kzarb} \end{equation}

We see that the result is $z$-independent, expressing the conservation of photon momentum along the direction of propagation.

\section{Physical Significance of the Gouy Radius}
\label{significance}
The (longitudinal) Gouy effect has been explained as a result of the transverse confinement or restriction of the photon beam~\cite{HariharanRobinson, FengWinful}. This restriction leads to the generation of transverse photon momentum or wave vector components, so that~\cite{FengWinful} 
\begin{equation} \langle k_z\rangle = k_0 -\frac{\langle k^2_x\rangle + \langle k^2_y\rangle}{k_0} \,,\label{transverse} \end{equation} where it is to be noted that $ \langle k_z\rangle^2 \neq \langle k^2_z\rangle $.
Comparing with Eq.~(\ref{kzarb}) we obtain
\begin{equation} \langle k^2_x\rangle + \langle k^2_y\rangle = \frac{2p+|\ell| +1}{(w(0))^2} \,, \end{equation}
where the average is performed over a plane of constant $z$,
or, using the cylindrical symmetry of the beam, \begin{equation} \langle k^2_x \rangle = \langle k^2_y\rangle = \frac{2p+|\ell| +1}{2 (w(0))^2} \,. \end{equation}
These equations express the position-momentum uncertainty in the transverse direction.

Now if we consider averaging over $\phi$ only in Eq.~(\ref{transverse}), at fixed $z$ and $r$, we see that within the Gouy radius the transverse wave numbers $k_x$ and $k_y$ are real, but beyond this radius $k_x$ and $k_y$ become imaginary. Imaginary wave numbers imply amplitude damping. The situation is analogous to a particle confined by a finite potential wall. There is a finite probability of penetration within the potential wall, and the probability drops off rapidly within this ``forbidden'' zone. Thus the Gouy radius provides a more satisfactory ``physical'' boundary for the photon beam than the function $w(z)$ or the related radius of maximum intensity or the rms radius.

We shall now offer a correction to an important result that was derived for the rotation of the Poynting Vector in the course of the propagation of a Laguerre Gaussian beam \cite{PadgAll}. The authors have obtained an expression for the winding of the Poynting Vector around the axis of propagation of the photon beam. Taking the Poynting Vector to be concentrated at the radius of maximum field they showed that this Poynting Vector undergoes a rotation of $\pi/2$ when propagating from the beam waist to the far field, and that the rotation is independent of the value of $\ell$.  But a more accurate calculation shows that the Poynting Vector does indeed circulate as a corkscrew and that the rotation is dependent on the value of $\ell$.

The Poynting vector $\vec{S}$ of an LG beam is obtained by taking the gradient of $u$ in Eq.~(\ref{LG}):
\begin{equation} 
\vec{S}  = \Xi(r,z)\left[\frac{zr}{z^2_R+z^2}\hat{r}+\frac{|\ell|}{k_0 r}\hat{\phi} +\left(1+\frac{r^2(z^2_R-z^2)}{2(z^2_R+z^2)^2} -\frac{(2p+|\ell| +1)z_R}{(z^2_R +z^2)k_0}\right)\hat{z}\right] \, \end{equation} where $\Xi$ is a function of $r$ and $z$. 
(The expression given in Ref.~\cite{PadgAll} is missing a square and a minus sign.)
In Ref.~\cite{PadgAll}, the second and third terms inside the parentheses are assumed to be small and neglected. But for $r$ of order of the beam width, $r^2 \sim z_R/k_0$ (Eq.~(\ref{Gauss})), these terms contribute to the Poynting vector magnitude $|\vec{S}|$ at the same order as the $\hat{\phi}$ term.

What is clear is that dropping these two terms is equivalent to setting their sum equal to zero, which is valid but {\it only} at the Gouy radius $r_G$. So in the expression given in Ref.~\cite{PadgAll} for the rate of azimuthal rotation,
\begin{equation} \frac{\partial \varphi}{\partial z} = \frac{|\ell|}{k_0 r^2}\,,\label{rotationrate} \end{equation}
$r$ is not an arbitrary radius but precisely the Gouy radius $r_G$. It is therefore problematic to assign to $r$ the value of the radius of maximum field intensity~\cite{PadgAll} or any radius other than the Gouy radius.

Substituting for $r_G$ from Eq.~(\ref{GRz1}), we obtain
\begin{equation} \frac{\partial \varphi}{\partial z}= \frac{|\ell|}{2z_R(2p+|\ell|+1)}\frac{(z^2_R -z^2)}{(z^2_R +z^2)}\,,\label{rateofrotation}\end{equation}
and upon integrating, we have
\begin{equation} \varphi =  \frac{|\ell|}{(2p+|\ell|+1)}\left[\tan^{-1}\left(\frac{z}{z_R}\right) - \frac{z}{2z_R}\right]\label{theta}\,.\end{equation}
Thus far, we have not made any approximations. Eq.~(\ref{theta}) follows exactly from Eq.~(\ref{LG}). But Eq.~(\ref{LG}) itself is an exact solution to Maxwell's equations only in the paraxial approximation, $z \ll z_R$. 
Eq.~(\ref{rateofrotation}) shows that at small distances from the focus, $z \ll z_R$, the rotation rate is
\begin{equation} \frac{\partial \varphi}{\partial z} = \frac{|\ell|}{2z_R(2p+|\ell|+1)}\,, \label{paraxialrate1}\end{equation} to be compared with the rate $\partial \varphi/ \partial z = 1/z_R$ obtained in
Ref.~\cite{PadgAll}.

\section{Conclusions}

We have analyzed the transverse structure of the Gouy effect in a Laguerre-Gaussian twisted photon beam, with an emphasis on the Gouy radius $r_G$, defined as the critical radius where the Gouy phase shift vanishes. We have shown that in the focal plane of the LG beam, the Gouy radius is proportional to the rms radius of the beam, which in turn scales as $\sqrt{2p+|\ell|+1}$ with the angular momentum $\ell$ and nodal quantum number $p$ of the beam. However, away from the focal plane, $r_G$ grows faster than the rms radius, and diverges at the Rayleigh range $z_R$, even while the beam width and the rms radius remain finite. The Gouy radius is shown to delineate a boundary between the ``allowed'' and ''forbidden'' zones of the photon beam, and is conducive to understanding the rotation of the Poynting vector in a twisted beam. 

\acknowledgments
We thank J. H. McGuire for helpful discussions. This work was supported in part by the NSF under Grant No. PHY-1005709.

\end{document}